# BLIND 3D MODEL WATERMARKING BASED ON MULTI-RESOLUTION REPRESENTATION AND FUZZY LOGIC

[1]Prof. Sharvari C. Tamane, [2]Dr. Ratnadeep R. Deshmukh,

[1]MGM's Jawaharlal Nehru Engineering College
sharvaree73@yahoo.com
[2]Dr. Babasaheb Ambedkar Marathwada University,
ratnadeep_deshmukh@yahoo.co.in

***ABSTRACT***

*Insertion of a text message, audio data or/and an image into another image or 3D model is called as a watermarking process. Watermarking has variety of applications like: Copyright Protection, Owner Identification, Copy Protection and Data Hiding etc., depending upon the type of watermark insertion algorithm. Watermark remains in the content after applying various attacks without any distortions. The blind watermarking method used in the system is based on a wavelet transform, a fuzzy inference system and a multi-resolution representation (MRR) of the 3d model. The watermark scrambled by Arnold Transform is embedded in the wavelet coefficients at third resolution level of the MRR. Fuzzy logic approach used in the method makes it to approximate the best possible gain with an accurate scaling factor so that the watermark remains invisible. The fuzzy input variables are computed for each wavelet coefficient in the 3D model. The output of the fuzzy system is a single value which is a perceptual value for each corresponding wavelet coefficient. Thus, the fuzzy perceptual mask combines all these non-linear variables to build a simple, easy to use HVS model. Results shows that the system is robust against affine transformations, smoothing, cropping and noise attacks.*

## 1. INTRODUCTION

Digital watermarking [1] has been considered a potential efficient solution for copyright protection of various multimedia contents. This technique carefully hides some secret information in the functional part of the cover content. Compared with cryptography, the digital watermarking technique is able to protect digital works (assets) after the transmission phase and the legal access. There exist different classifications of watermarking algorithms. We distinguish between non-blind and blind watermarking schemes depending on whether or not the original digital work is needed at extraction phase resp. Watermark robustness is the ability to recover the watermark even if the watermarked 3D model has been manipulated. Usually, one hopes to construct a robust watermark which is able to go through common malicious attacks for copyright protection purposes. However, sometimes the watermark is intentionally designed to be fragile, even to very slight modifications, in order to be used in authentication applications. Watermarking algorithms are divided into two categories, spatial domain based or transform domain based, according to the insertion space. Spatial domain method embeds the watermark by directly modifying the original image data (pixel value), while in the frequency domain method, a watermark is embedded into coefficients by modifying it after taking the transform such as Discrete Cosine transform (DCT), Fast Fourier transform (FFT) and wavelet transform (WT).

Nowadays, 3D meshes [2] are widely used in virtual reality, medical imaging, video games and computer aided design. A mesh is a collection of polygonal facets targeting to constitute an appropriate approximation of a real 3D object. It has three different combinatorial elements: vertices, edges, and facets. From another viewpoint, a mesh can also be completely described by two kinds of information: the geometry information describes the 3D positions (coordinates) of all its vertices, while the connectivity information provides the adjacency relations between the different elements.

The method presented here is a combination of two algorithms proposed by Saeed K. Amirgholipour [3] and Mukesh Motwani [4] with some modifications. Paper [3] represented a novel blind watermarking algorithm based on a joint DWT-DCT for 2D digital image. He exploited strength of two common frequency domain methods; DCT and DWT, to obtain imperceptibility and robustness. The idea of inserting watermark in the combined transform was based on the fact that, the joint transform can eliminate the drawback of each other and then, an effective watermark is embedded in the most robust and imperceptible parts of the image. Watermarking is done with embedding the watermark in the special middle frequency coefficient sets of 3-levels DWT transformed of a host image, followed by computing 4×4 block-based DCT on the selected DWT coefficient sets. Paper [4] has proposed a nonblind watermarking algorithm based on wavelets and fuzzy logic, which inserts 8 bit grey scale image as a watermark into the 3D model.

The algorithm used in this paper embeds the watermark by modifying the wavelet coefficients of the 3 D model at third level and by using Mamdani's fuzzy inference system, thus enhancing the robustness and making it more imperceptible as compared with other existing algorithms. Particular attention would be paid to ensure that the embedding algorithm preserves the visual integrity of the models.

## 2. BACKGROUND

Wavelets [5] are used in various applications because it has scale and time aspects. Wavelet analysis can provide a windowing technique with varying length area. Some of the important advantages of wavelet analysis are as follows:

- It allows the use of variable size intervals (shorter or longer) to get high and low frequency information.
- It allows to function local analysis of a signal.
- It can compress or de-noise a signal without losing its quality.
- It also has advantages over other signals like decomposing of a signal into various parts and break-in-continuities for higher derivatives.

Wavelet analysis process divides the signal into two components: approximation and detail. Approximation and detail components are also called as low frequency and high frequency components resp. The decomposition of a signal into low and high frequency components is also called as analysis. The reconstruction of these components back into the original signal without loss of information is called synthesis. In Matlab decomposition process is performed by DWT function and the reconstruction process is performed by IDWT function.

Multi-resolution analysis [6] is a useful tool to reach an acceptable trade-off between the mesh complexity and the capacity of the available resources. Such an analysis produces a coarse mesh which represents the basic shape (low frequencies) and a set of details information at different resolution levels (median and high frequencies). These methods also allow realizing a synthesis process during which multiple representations with different complexities can be created.

The most interesting point of multi-resolution analysis for watermarking is its flexibility. There are different available locations allowing to meet different application demands. For example, insertion in the coarsest mesh ensures a good robustness, while embedding in the details parts provides an excellent capacity. The insertion in low resolution can be both more robust and more imperceptible thanks to a dilution effect. The insertion in high resolution level may permit to construct some effective fragile watermarks with a precise localization ability of the attacks.

Wavelets are a common tool for such a multi-resolution analysis. The watermark can be inserted either in the coarsest mesh, or in the wavelet coefficients at different levels. In fact, these wavelet coefficients are 3D vectors associated with each edge of the corresponding coarser mesh. Note that this kind of wavelet analysis is applicable only on semi-regular triangular meshes. Based on this wavelet analysis, Kanai et al. [7] proposed a non-blind algorithm that modifies the ratio between a wavelet coefficient norm and the length of its support edge, which is invariant to similarity transformations. Uccheddu et al. [8] described a blind one-bit watermarking algorithm with the hypothesis of the statistical independence between the wavelet coefficients norms and the inserted watermark bit string. Paper [9] represented a blind watermarking algorithm based on DWT-DCT coefficients for 3D surface models.

Fuzzy Inference system (FIS) [5] provides mapping of input to output using fuzzy logic. Output values are then used to take decisions or to classify patterns. FIS consists of membership functions, operators and If Then rules. Some important points about fuzzy logic are as follows:

- Fuzzy logic is conceptually easy to understand.
- Fuzzy logic is flexible.
- Fuzzy logic is tolerant of imprecise data.
- Fuzzy logic can model nonlinear functions of arbitrary complexity.
- Fuzzy logic can be easily mixed with Traditional techniques.
- Fuzzy logic is based on natural language.

Amongst the two types available in the Matlab fuzzy logic toolbox, Mamdani-type FIS is used in the paper. It is used to calculate the weight factor which helps to take decision that where to insert the watermark data in the 3D model? FIS consists of five steps:

Step 1: Fuzzification of the input parameters: All the input parameters (crisp values) are converted into fuzzy values (between zero and one) via membership functions.

Step 2: Application of fuzzy operators: Apply fuzzy operator (AND or OR) on input parameters and generate output as a truth value. The input parameter may contain more than two membership values. The output value is computed either by AND method (i.e. minimum and product method) or by OR method (i.e. maximum and probabilistic OR method).

Step 3: Application implication method: All the rules defined in the FIS have weight values in between zero and one. This weight value is applied on the output of the above step. Most of the times this weight value considered are one.

Step 4: Aggregation of all outputs: Aggregate all the outputs of rules into a fuzzy set to take a final decision. Aggregation is applied separately on each output parameter only once.

Step 5: Defuzzification: Fuzzy set can be converted into a crisp number. In Matlab toolbox five different methods are available to perform defuzzification on fuzzy set. These are: centroid, bisector, MOM (middle of maximum), LOM (largest of maximum), and SOM (smallest of maximum). Generally centroid method is the choice for defuzzification as it returns the center value.

Arnold’s transformation is a technique to scramble the watermark to increase the robustness of a model against cropping. It disorders the image matrix and makes the image obscured. This increases the watermark secrecy and security. We can get back the original watermark image by applying Arnold transform on the scrambled image. Arnold transform equation is shown below:

$$\begin{Bmatrix} P' \\ Q' \end{Bmatrix} = \left( \begin{pmatrix} 1 & 1 \\ 1 & 2 \end{pmatrix} \begin{Bmatrix} P \\ Q \end{Bmatrix} \bmod N \right)$$

Where, P and Q denote the coordinate of pixels of the original watermark, P’ and Q’ denote the coordinate of pixels of the transformed watermark, and N denotes the size of the watermark.

## 3. WATERMARKING FOR A 3D MODEL

3d models used here consists of three matrices (x1, x2, x3) of dimension N * N (N being the size of matrix) for x, y and z direction respectively.

The basic steps for insertion of watermark are explained in following subsection. First 09 steps are applied on X direction matrix:

Step 1: Apply Haar wavelet transform on the 3d model and generate four coefficient sets as approximation coefficients matrix CA and details coefficients matrices as CH, CV and CD.

Step 2: Apply Haar wavelet transform again on CH and CV of detail coefficient matrices to generate level 2 approximation and detail coefficient matrices. Total 8 matrices are generated in this step.

Step 3: Apply Haar wavelet transform again on CH and CV coefficients of level 2 matrices to generate level 3 approximations and detail coefficients. Total 16 matrices are generated in this step.

Step 4: Convert the watermark image into binary format.

Step 5: Scramble the watermark with Arnold transform for key times.

Step 6: Compute Inputs (like Curvature, Area and Bumpiness) of 3d model.

Step 7: Apply inputs to Fuzzy Inference system (FIS) and decide weight depending upon rules provided in the FIS for each input values. The fuzzy output has 7 membership functions (like Lowest, Lower, Low, Medium, High, Higher, Highest), only High and HIGHER fuzzy output sets are used for insertion of watermark in the 3D model. This is to make the watermark imperceptible and more robust.

A total of 15 fuzzy rules are developed, one of the rules is as follows:

Rule: IF Curvature == MEDIUM and

Bumpiness == MEDIUM and Area == LOW

THEN Weighting factor = LOW.

Step 8: Modify coefficients to embed the watermark into the model where the corresponding weighting factor has the values either high or higher. The watermark is inserted by adjusting the remainder of wavelet coefficient vector greater than i if watermark bit is 1 or less than j if watermark bit is 0 in such a way that it should not disturb the visibility(or quality) of the model, where i and j are some constants and i>j.

Step 9: Apply inverse DWT on the modified coefficients set up to level 3.

Step 10: Repeat steps 1, 2, 3, 8 and 9 for Y direction matrices to generate watermarked 3d model.We can also insert watermark data in z direction matrix to increase the watermark insertion capacity.

## 4. WATERMARK EXTRACTION ALGORITHM

Watermark can be extracted from the 3D watermarked model by using a blind method. Algorithm for extracting the watermark is as follows:

Step 1: Perform the first three steps of insertion process on 3d watermarked model.

Step 2: Use the same FIS as used in the embedding process and calculate the remainder of the coefficients for high and higher weight values of FIS. If the remainder of the coefficient is greater than i then the extracted watermark bit is 1 otherwise it is 0.

Step 3: The scrambled watermark is reconstructed using the extracted watermark bits with Arnold transform and calculate the correlation between original and extracted watermark.

Correlation coefficient is used to evaluate the class of the extracted watermark by computing the similarity of the original watermark (W) and the extracted watermark (W’). The value of correlation is between zero and one. A larger value of correlation represents more similarity of the original watermark and extracted watermark. In Matlab Correlation between two matrices can be calculated by using a function corr2 as: r = corr2 (A, B)

Where r is the correlation coefficient between A & B, and A & B are matrices or vectors (original and extracted watermarks resp.) of the same size. Formula for the corr2 function is…

$r = (\sum_m \sum_n (A_{mn}-A') (B_{mn}-B'))/(sqrt(\sum_m \sum_n (A_{mn}-A')^2 ) (\sum_m \sum_n (B_{mn}-B')^2))$.

Peak signal to noise PSNR ratio is the most important technique to verify the quality of watermarked model. Clearly, the larger the payload, the lower the PSNR will be. The large payload means the strong embedding strength. This is generally true for all data hiding methods. Tables 2, 3 & 4 show the performance in terms of Correlation coefficient and PSNR.

## 5. ATTACKS

By applying rotation and translation transformation on 3d Watermarked model, only position of model is changed and therefore it is invariant to translation and rotation transformation.

By applying scaling transformation on 3d watermarked model, magnitude of wavelet coefficients are changed, but due to normalization of model during insertion and extraction process the watermark is unaffected.

Random noise of 10% amplitude is added to the 3d watermarked model because noise more than that may destroy watermark present in the model. The model is also attacked by salt and pepper noise with .05 and .1 noise density, and still it can detect watermark.

To perform smoothing attack on 3d watermarked model Laplacian, Gaussian and LoG Matlab Smoothing filters are used. These filters may reduce the quality of the 3d model, but the extracted watermark can still represent the correct watermark. Smoothing parameters values are given in the table.

Cropping of 9% and 16% is applied on the watermarked model but it has not destroyed the watermark completely because watermark is inserted uniformly at the embedding process in the model.

Thus the system has been proved to be robust against smoothing, similar transformations (Rotation, Translation and Scaling), noise and cropping attacks.

## 6. EXPERIMENTAL RESULTS

Robust and blind algorithms are very attractive because of their reliability and flexibility as inserting watermark at different locations allows to meet different application demands in terms of robustness and capacity. The effectiveness of the system is tested for the models given in a Table 1. Text, 2d image or audio data can be considered as a watermark. Size of the watermark depends upon the size of the 3d model. In the implementation key value considered is 5 and the values for i & j are calculated on trial and error basis. From the Tables 2, 3 & 4 it can be seen that the embedded watermark is imperceptible and robust to noise, smoothing and cropping attacks.

**Table 1. 3D Models and watermark**

| | |
|---|---|
| 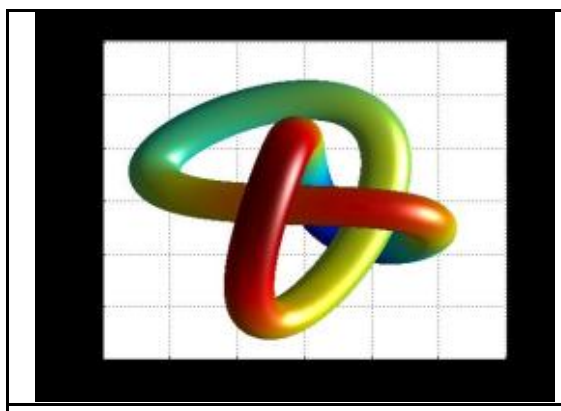 | 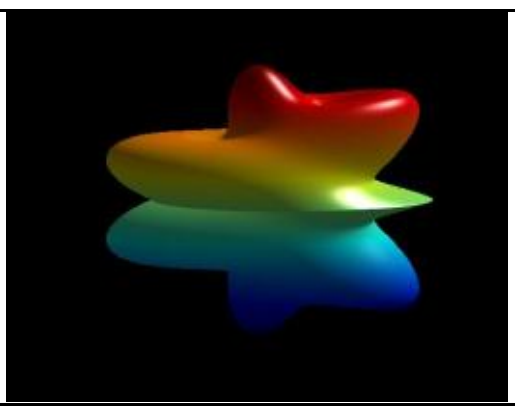 |
| 3D model 1 | 3D Model 2 (Harmonic) |
| 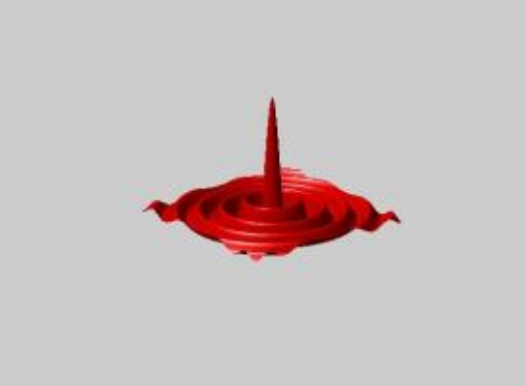 | 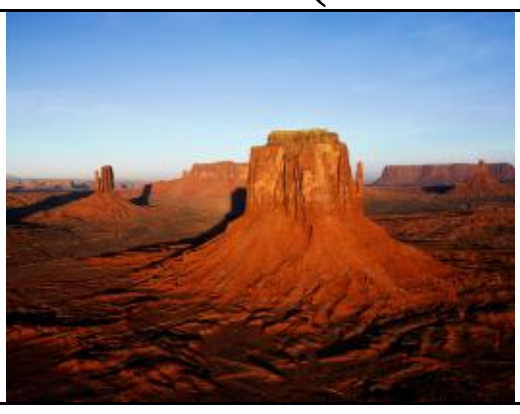 |
| 3D Model 3 (Mesh) | Watermark |

**Table 2. Results of Model 1**

| Attack | Correlation (of original watermark and extracted watermark ) | PSNR (of original model & watermarked model) | Extracted Watermark |
|---|---|---|---|
| Extracted Watermark Without Any Attack | 1.000000 | 92.450336 | 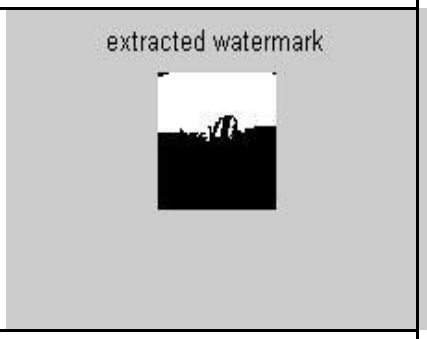 |
| **Gaussian Smoothing** | | | |
| hsize=3, sigma=10 | 0.471987 | 92.450336 | 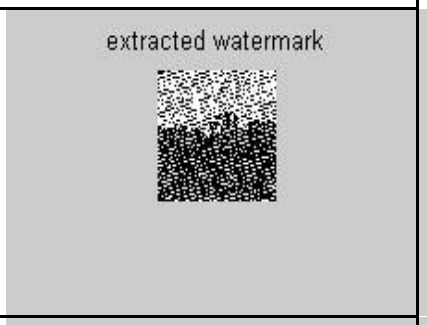 |
| hsize=7, sigma=10 | 0.358508 | 92.450336 | 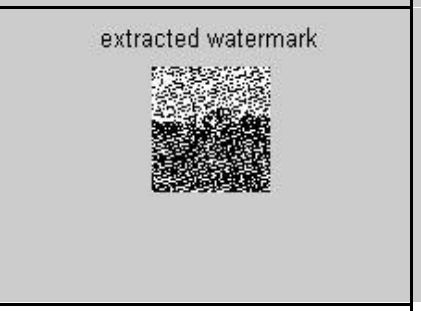 |
| **Laplacian Smoothing** | | | |
| Alpha=1 | 0.675534 | 92.450336 | 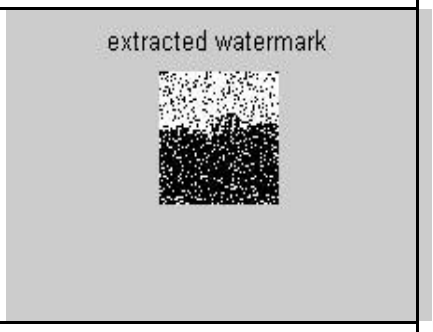 |
| **Log Smoothing** | | | |
| hsize=5, sigma=0.5 | 0.421914 | 92.450336 | 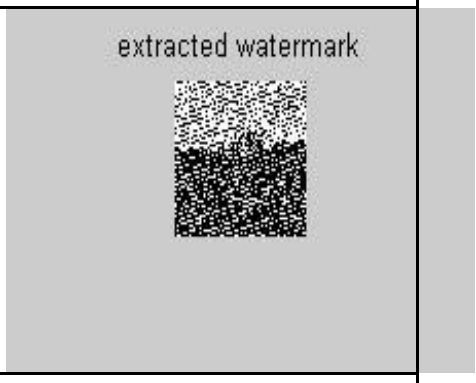 |
| **Salt & Pepper Noise** | | | |
| D=5 % | 0.796418 | 92.450336 | 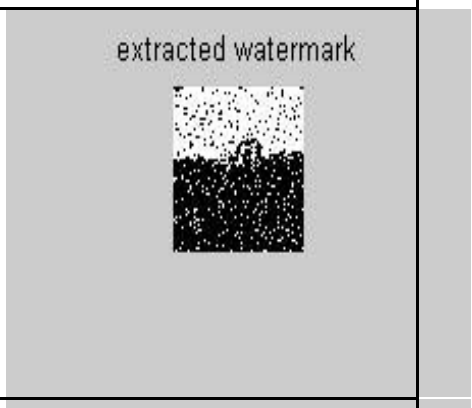 |
| D=10 % | 0.721926 | 92.450336 | 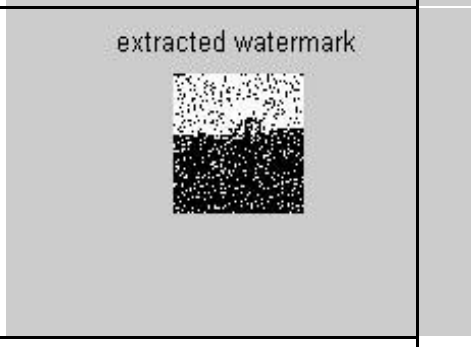 |
| **Random Noise** | | | |
| 0.1 | 0.693168 | 86.693808 | 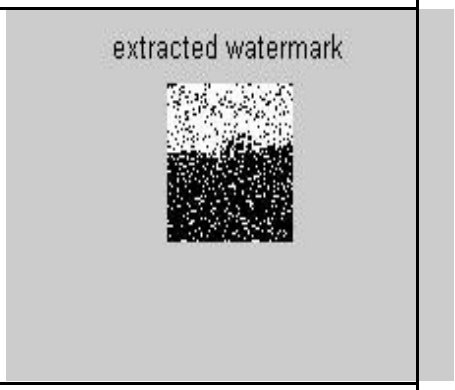 |
| **Cropping** | | | |
| 9 % | 0.992457 | 92.450336 | 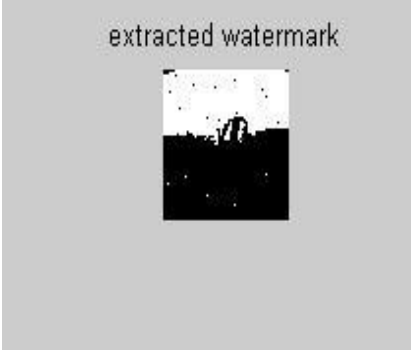  |
| 16 % | 0.979363 | 92.450336 | 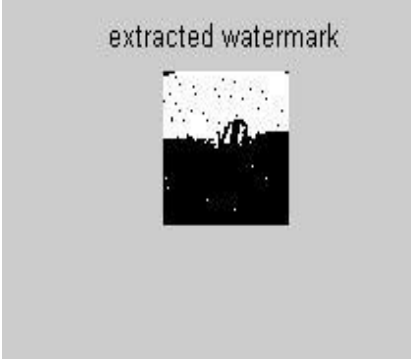  |

**Table 3. Results of model 2(Harmonic)**

| Attack | Correlation (of original watermark and extracted watermark ) | PSNR (of original model & watermarked model) | Extracted Watermark |
|---|---|---|---|
| Extracted Watermark Without Any Attack | 1.000000 | 74.784541 | 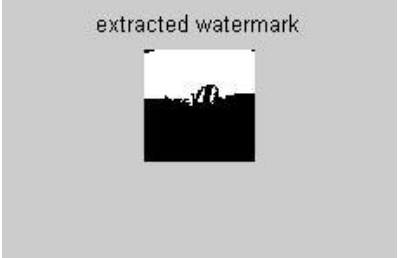  |
| **Gaussian Smoothing** | | | |
| hsize=3, sigma=10 | 0.448048 | 74.784541 | 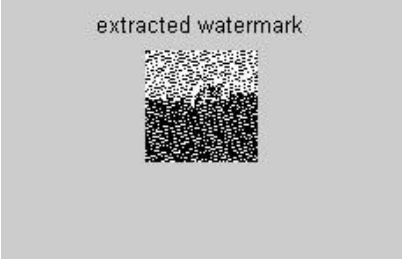  |
| hsize=7, sigma=10 | 0.333929 | 74.784541 | 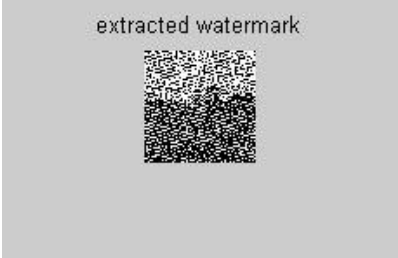  |
| **Laplacian Smoothing** | | | |
| Alpha=1 | 0.676401 | 74.784541 | 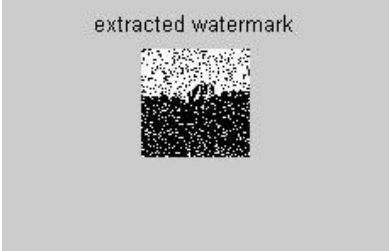  |
| **Log Smoothing** | | | |
| hsize=5, sigma=0.5 | 0.409964 | 74.784541 | 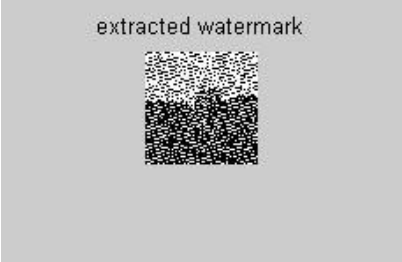  |
| **Salt & Pepper Noise** | | | |
| D=5 % | 0.424623 | 74.784541 | 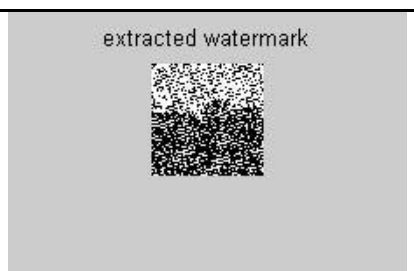  |
| D=10 % | 0.359999 | 74.784541 | 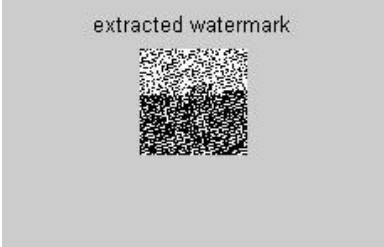  |
| **Random Noise** | | | |
| 0.1 | 0.377046 | 38.687482 | 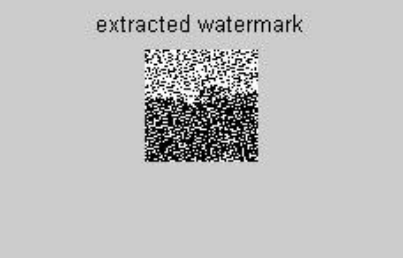  |
| **Cropping** | | | |
| 9 % | 0.991437 | 74.784541 | 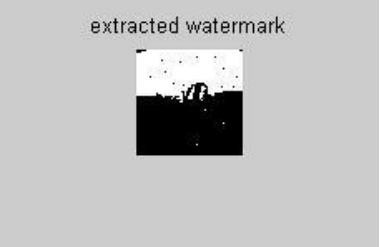 |
| 16 % | 0.981384 | 74.784541 | 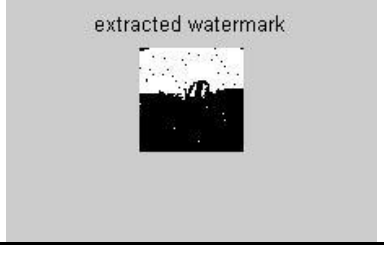 |

**Table 4.Results of Model 3 (Mesh)**

| Attack | Correlation (of original watermark and extracted watermark ) | PSNR (of original model & watermarked model) | Extracted Watermark |
|---|---|---|---|
| Extracted Watermark Without Any Attack | 0.934571 | 90.203495 | 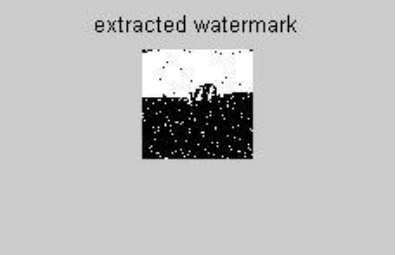 |
| **Gaussian Smoothing** | | | |
| hsize=3, sigma=10 | 0.377116 | 90.203495 | 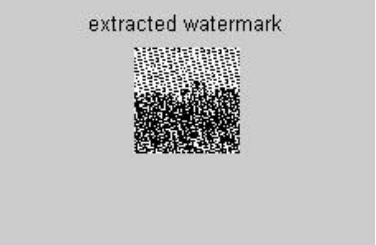 |
| hsize=7, sigma=10 | 0.219841 | 90.203495 | 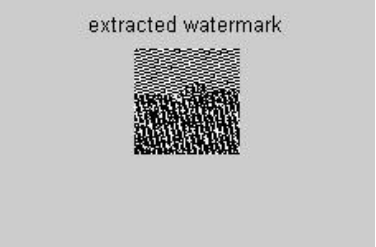 |
| **Laplacian Smoothing** | | | |
| Alpha=1 | 0.632348 | 90.203495 | 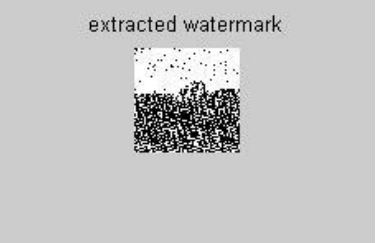 |
| **Log Smoothing** | | | |
| hsize=5, sigma=0.5 | 0.385202 | 90.203495 | 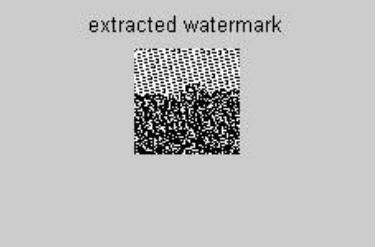 |
| **Salt & Pepper Noise** | | | |
| D=5 % | 0.455987 | 90.203495 | 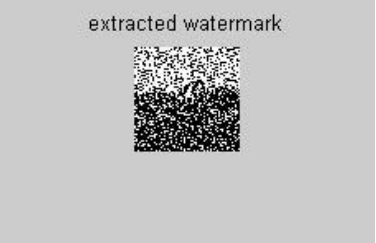 |
| D=10 % | 0.402706 | 90.203495 | 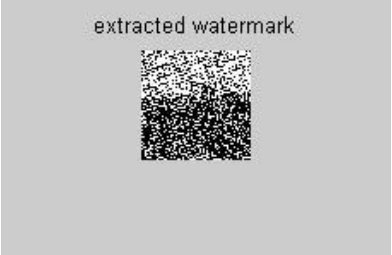 |
| **Random Noise** | | | |
| 0.1 | 0.337013 | 52.896630 | 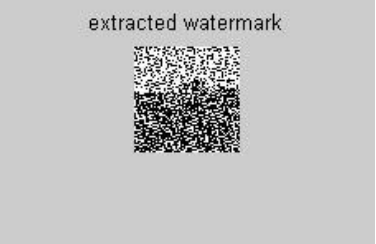 |
| **Cropping** | | | |
| 9 % | 0.926507 | 90.203495 | 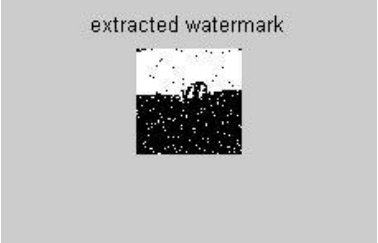 |
| 16 % | 0.917646 | 90.203495 | 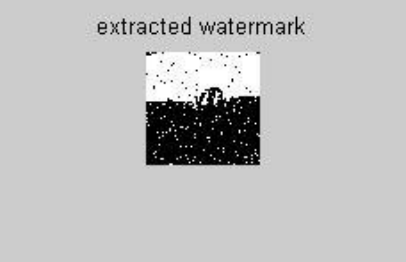 |

## 7. CONCLUSIONS & FUTURE SCOPE

We have evaluated a novel, blind & robust watermarking algorithm that embeds a watermark by modifying wavelet coefficient at level three considering weight factor calculated by FIS. Our system is able to distribute watermark over the entire model. Experimental result shows that this approach is able to withstand common attacks like affine transformation, smoothing, noise and cropping. Future watermark embedding algorithms can be developed which should be invariant to all types of attacks and will not change the imperceptibility.

## ACKNOWLEDGEMENT

We would like to thank to the authors of the papers [3] and [4] whose algorithms motivated us to implement the algorithm for 3d regular models with some modifications.